\documentclass[a4paper,11pt]{article}
\pdfoutput=1 % if your are submitting a pdflatex (i.e. if you have
\usepackage[utf8]{inputenc}
\usepackage{amsmath}
\usepackage{amsthm}
\usepackage{amssymb}
\usepackage[top=1in, bottom=1in, left=1in, right=1in]{geometry}
\usepackage{booktabs}
\usepackage[english]{babel}
\usepackage{color,soul}
\usepackage{graphicx}
\usepackage{caption}
\usepackage{subcaption}
\usepackage{wrapfig}
\usepackage{physics}
\usepackage{slashed}
\usepackage{breqn}
\usepackage{cancel}
\usepackage{mathtools}

\DeclarePairedDelimiter\floor{\lfloor}{\rfloor}
\usepackage[OT2,T1]{fontenc}
\DeclareSymbolFont{cyrletters}{OT2}{wncyr}{m}{n}
\DeclareMathSymbol{\Sha}{\mathalpha}{cyrletters}{"58}

\usepackage{tikz}
\usetikzlibrary{backgrounds,shapes,arrows,positioning,calc,snakes,fit}
\usepgflibrary{decorations.markings}

\usepackage{color}   %May be necessary if you want to color links

\usepackage{jheppub} 

\usepackage[T1]{fontenc} % if needed

\title{\boldmath Resurgent analysis of SU(2) Chern-Simons partition function on Brieskorn spheres $\Sigma(2,3,6n+5)$}

\author{David H. Wu}
\affiliation{Department of Physics, California Institute of Technology,\\Pasadena, CA 91125 USA}
\abstract{$\hat{Z}$-invariants, which can reconstruct the analytic continuation of the SU(2) Chern-Simons partition functions via Borel resummation, were discovered by GPV and have been conjectured to be a new homological invariant of 3-manifolds which can shed light onto the superconformal and topologically twisted index of 3d $\mathcal{N}=2$ theories proposed by GPPV. In particular, the resurgent analysis of $\hat{Z}$ has been fruitful in discovering analytic properties of the WRT invariants. The resurgent analysis of these $\hat{Z}$-invariants has been performed for the cases of $\Sigma(2,3,5),\ \Sigma(2,3,7)$ by GMP, $\Sigma(2,5,7)$ by Chun, and, more recently, some additional Seifert manifolds by Chung and Kucharski, independently. In this paper, we extend and generalize the resurgent analysis of $\hat{Z}$ on a family of Brieskorn homology spheres $\Sigma(2,3,6n+5)$ where $n\in\mathbb{Z}_+$ and $6n+5$ is a prime. By deriving $\hat{Z}$ for $\Sigma(2,3,6n+5)$ according to GPPV and Hikami, we provide a formula where one can quickly compute the non-perturbative contributions to the full analytic continuation of SU(2) Chern-Simons partition function.}

\begin{document}

\maketitle
\flushbottom

\section{Introduction} 
Consider the Chern-Simons theory defined on a 3-manifold $M_3$ with a gauge group G that has a corresponding Lie algebra g and A as the one-form gauge connection on $M_3$ in g:
\begin{equation}
\label{eq:chernsimonsFunc}
    \text{CS}(A)=\frac{1}{8\pi^2}\int_{M_3}A\wedge dA+\frac{2}{3}A\wedge A\wedge A.
\end{equation}
An interesting quantity that captures the physics on the 3-manifold is the Feynman path integral which is typically defined as 
\begin{equation}
\label{eq:partFuncCS}
    Z(M_3)=\int_{\mathcal{A}_{SU(2)}} DA e^{2\pi ik\text{CS}(A)}
\end{equation}
where k is the coupling constant, $\mathcal{A}_{SU(2)}$ is the space of gauge connections on $M_3$, and we work in the Euclidean signature \cite{GMP}.\footnote{For the following sections, we identify
\begin{equation}
    q=e^{\hbar}=e^{2\pi i/k}=e^{2\pi i\tau}
\end{equation}
and will use these variables interchangeably similar to that done in the literature \cite{GMP,GS} for Chern-Simons theory.} The Chern-Simons theory consequently yields a set of equation of motion s.t. the classical solution to the theory consists of the flat connections satisfying the EOM. Additionally, one could compute the partition function as a sum of equally weighted perturbative contributions from saddle points, $e^{-\frac{1}{\hbar}\text{CS}(\alpha)}Z_{\alpha}(\hbar)$ where $\alpha$ denotes the saddle points in the theory. However, this perturbation method will fail in its analytic continuation when $\hbar$ becomes large or acquires a complex phase. Therefore, alternatively, the exact partition function for the Chern-Simons theory on a 3-manifold can be derived using resurgent methods introduced in \cite{Ecalle}
\begin{equation}
\label{eq:EcalleTrans}
    Z(\hbar)=\sum_{\alpha}n_{\alpha}e^{-\frac{1}{\hbar}\text{CS}_{\alpha}}Z^{\alpha}_{\text{pert}}(\hbar)
\end{equation}
where $n_{\alpha}$ are the transseries parameters of which accounts for the contribution from each saddle point to the analytically continued Feynman path integral. 

%To be more concrete on our approach, we start with Eq. \eqref{eq:partFuncCS} integrated over the SU(2) connections as specified while analytically continuing the coupling constant to the complex plane. Now, we can utilize well-established methods in standard QFT to perform the integral of steepest descent \cite{Marino,GLM,Garoufalidis,Witten}. Given that our main subject of interest is the non-perturbative behavior in Eq. \eqref{eq:partFuncCS}, we now notice that the integral can be conveniently perturbed and expressed as a trans-series \cite{GMP}. Then it is natural to perform a Borel resummation of the series to obtain information on additional physics. To build up our intuition, in the next two subsections, we will provide a brief introduction to the techniques of Borel resummation and elaborate on the ideas illustrated in this paragraph.

To understand and motivate the definition of Eq. \eqref{eq:EcalleTrans}, we start from Eq. \eqref{eq:partFuncCS} and extend the coupling constant k to complex values where the usual perturbation method fails. Established in \cite{Pham}, we aim to utilize the precursors to resurgent analysis, Picard-Lefschetz and associated Stokes phenomena, to solve for the partition function. (Illuminating implementations in the context of exponential integrals can be found in \cite{Witten,Garoufalidis,Kontsevich}.) In complexifying the coupling constant k and performing an integration by the steepest descent integration cycles, we have 
\begin{equation}
    Z_{\text{CS}}(M_3)=\int_{\Gamma_{SU(2)}\subset \tilde{A}_{SL(2,\mathbb{C})}} DAe^{2\pi i k\text{CS}(A)}
\end{equation}
where $\tilde{A}_{SL(2,\mathbb{C})}$ is now the universal cover of the $SL(2,\mathbb{C})$ flat connections modulo gauge equivalence. In other words \cite{Chun}, we are integrating over a middle-dimension cycle $\Gamma_{SU(2)}$ of which lives in the moduli space of $SL(2,\mathbb{C})$ flat connections. Now we can organize our integration domains by their corresponding instanton number:
\begin{equation}
\label{eq:LThimbles}
    \Gamma=\sum_{\tilde{\alpha}\in\pi_0(\mathcal{M}_{\text{flat}}(M_3,SL(2,\mathbb{C})))\times \mathbb{Z}}n_{\tilde{\alpha},\theta}\Gamma_{\tilde{\alpha},\theta}
\end{equation}
of which $\Gamma_{\tilde{\alpha},\theta}$ is the union of steepest descent flows from the saddle points 
\begin{equation}
\label{eq:saddles}
    \tilde{\alpha}\in \tilde{M}_{\tilde{\alpha}}\subset \text{Hom}(\pi_1(M_3),SL(2,\mathbb{C}))\times \mathbb{Z}
\end{equation} 
and $n_{\tilde{\alpha},\theta}\in\mathbb{Z}$ are the transseries parameters, and $\theta=\arg{k}$ is a general value dictating the direction of Borel resummation in the Borel plane.

Now, as we successfully decomposed our interested integration cycles into sums over Lefschetz thimbles, by linearity of integration, we can now focus simply on individual Lefschetz thimbles at a time, distinguished by the instanton number, $\tilde{\alpha}$ and the general parameter $\theta$:
\begin{equation}
\label{eq:integralPart}
    I_{\tilde{\alpha},\theta}=\int_{\Gamma_{\tilde{\alpha},\theta}} DAe^{2\pi ik (\text{CS}(\alpha)+m)}
\end{equation}
where $m\in\mathbb{Z}$ and accounts for the value of the CS action prior to the modulation of 1 reflected in the CS invariant's numerical value. 
%Note that this gives us our transseries expansion of Eq. \eqref{eq:partFuncCS}:
%\begin{align}
%\label{eq:transseriesPartCS}
%\begin{split}
%    Z_{\text{CS}}(M_3;k)&=\sum_{\tilde{\alpha}\in\pi_0(\mathcal{M}_{\text{flat}}(M_3,SL(2,\mathbb{C})))\times \mathbb{Z}} n_{\tilde{\alpha},\theta}I_{\tilde{\alpha},\theta}\sim \sum_{\tilde{\alpha}} n_{\tilde{\alpha},\theta}e^{2\pi i k (CS(\alpha)+m)}Z^{\tilde{\alpha}}_{\text{pert}}(k)\\
%    &=\sum_{\tilde{\alpha}\in\pi_0(\mathcal{M}_{\text{flat}}(M_3,SL(2,\mathbb{C})))\times \mathbb{Z}} n_{\tilde{\alpha},\theta}\cdot{}\sum_{n=0}^{\infty} a_n^{\alpha}k^{-n+d_{\alpha}/2-3/2}
%    \end{split}
%\end{align}
%via Eq. \eqref{eq:LThimbles} where we obtained the second line from the asymptotic expansion w.r.t. $1/k$ of $I_{\tilde{\alpha},\theta}$ and more directly, we get 
%\begin{equation}
%    Z_{\text{pert}}^{\alpha}\in k^{(d_{\alpha}-3)/2}\mathbb{C}[[1/k]]
%\end{equation}
%where $d_{\alpha}=dim_{\mathbb{C}}\tilde{M}_{\tilde{\alpha}}$ from Eq. \eqref{eq:saddles} and $\alpha\in \pi_0(\mathcal{M}_{\text{flat}}(M_3,SL(2,\mathbb{C})))$ is the equivalence class of $\tilde{\alpha}$.
As $\theta$ is a parameter introduced in the general Borel resummation \cite{Marino}, for particular values of $\theta$, Eq. \eqref{eq:integralPart} will experience discontinuities/jumps known as the Stokes phenomenon. To account for the Stokes phenomenon occurring while we vary $\theta$, the integral then takes on the form
\begin{equation}
    I_{\tilde{\alpha},\theta}\bigg\rvert_{\theta=\theta_{\tilde{\alpha}\tilde{\beta}}+\epsilon}=I_{\tilde{\alpha},\theta}\bigg\rvert_{\theta=\theta_{\tilde{\alpha}\tilde{\beta}}-\epsilon}+m_{\tilde{\alpha}}^{\tilde{\beta}}I_{\tilde{\beta},\theta}\bigg\rvert_{\theta=\theta_{\tilde{\alpha}\tilde{\beta}}+\epsilon}
\end{equation}
where $\theta_{\tilde{\alpha}\tilde{\beta}}\equiv\arg\left(\text{CS}(\alpha)-\text{CS}(\beta)+\mathbb{Z}\right)/i$, $m_{\tilde{\alpha}}^{\tilde{\beta}}\in\mathbb{Z}$
is the Stokes monodromy coefficients\footnote{For more detailed and interesting discussions on the topic, see \cite{Witten,Kontsevich,Kontsevich2,GMN}. In particular, Fig. 9 in \cite{GMP} illustrates visually what the Stokes phenomenon is in the Borel plane.}, and $|\epsilon|\ll 1$ when such occurs for specific values of $\theta$. Equipped with these decomposed segments of the partition function, we can recover the full Chern-Simons partition function via Borel resummation \cite{GMP,Marino}. Due to the tower of solutions for the analytic continuation of Chern-Simons partition function, we can fix our attention to the case of $\theta=0$ as we can then obtain the reduction of our transseries in Eq. \eqref{eq:EcalleTrans}:
\begin{equation}
\label{eq:decompCSfull}
    Z_{\text{CS}}=\sum_{\alpha\in \pi_0(\mathcal{M}_{\text{flat}}(M_3,SL(2,\mathbb{C})))}n_{\alpha}e^{2\pi i\text{CS}(\alpha)}Z_{\text{pert}}^{\alpha}(k)
\end{equation}
where $n_{\alpha}\equiv \sum_{\tilde{\alpha}(\text{fixed }\alpha)}n_{\tilde{\alpha},\theta=0}$. With this identification, we can categorize the flat connections characterized by their corresponding stabilizers on $\text{Hom}(\pi_1(M_3),SU(2))$ employed in the context of \cite{GMP}: (1) SU(2) - central; (2) $U(1)$ - abelian; and (3) $\{\pm 1\}$ - irreducible. 

The existence of such $Z_{\text{pert}}^{\alpha}$, which is also denoted as $\hat{Z}_{\alpha}$ where $\alpha$ is an abelian flat connection in the theory, has been proposed in \cite{GPPV} and is conjectured to be a new homological 3-manifold invariant which admits integer powers and integer coefficients and is purely motivated by physics \cite{GPV}. This quantity is hence referred to as the homological block that can reconstruct the full SU(2) Chern-Simons partition function and can be computed from the graded Euler characteristics of the massless multiparticle BPS spectrum of the 3d $\mathcal{N}=2$ theory $T[M_3]$ with $\alpha \in \text{Tor}H_1(M_3,\mathbb{Z})/\mathbb{Z}_2$ as the discrete charge, denoted as $\mathcal{H}_{\alpha}$\footnote{The label $\alpha$ can be understood as giving rise to physical boundary conditions that come to realize BPS particles of $T[M_3]$ in M2-branes ending on a pair of M5-branes which realize the $A_1$ 6d $\mathcal{N}=(2,0)$ theory\cite{GPPV}.}\cite{GPPV,GPV}. This in turns can be computed as a SUSY partition function of $T[M_3]$ on $D^2\times_qS^1$ with the physical boundary set by $\alpha$:
\begin{equation}
    \hat{Z}_{\alpha}(q)=Z_{T[M_3]}(D^2\times_qS^1;q).
\end{equation}
While knowing the integrability of the physical BPS Hilbert space, $\mathcal{H}_{\alpha}$, resurgent analysis on the homological blocks can reveal the special roles abelian flat connections play in the integrability of q-series 3-manifold invariants and modularity of 3d $\mathcal{N}=2$ theories \cite{CCFGH}. These homological blocks, $\hat{Z}_{\alpha}$, or $Z_{\text{pert}}^{\alpha}$, will be the main physical quantity of interest as we perform the resurgent analysis of the SU(2) Chern-Simons partition function on a family of Brieskorn spheres $\Sigma(2,3,6n+5)$ with $n\in \mathbb{Z}_+$ and $6n+5$ is prime.

In this paper, the main goal is to extend the resurgent analysis of the analytically continued SU(2) Chern-Simons partition function defined on a family of Brieskorn spheres $\Sigma(2,3,6n+5)$ where $n\in \mathbb{Z}_+$ and $6n+5$ is prime. We provide supporting evidence to the claims made in \cite{GMP} on the $\hat{Z}$-invariants introduced in \cite{GPPV} that has already been initiated in various contexts \cite{Chun,Kucharski,Chung,Chung2019,Chung2018,CGPS}\footnote{A brief discussion on the resurgent analysis of $\hat{Z}$ on general Brieskorn spheres was presented in Sec. 2.6 of \cite{Chung2018}}. Similar to the discussion in Sec. 3.4 and 3.5 in \cite{GMP} and \cite{Chun}, We will start with, in section \ref{sec:2}, the general partition function for our interested Brieskorn spheres $\Sigma(2,3,6n+5)$ from the generating function introduced in \cite{GPPV} along with special cases of \cite{Hikami,Chung2018}. Then, using the resurgence properties of mock modular forms to analyze the homological blocks \cite{GMP}, we will provide a general expression of the Borel transformed Chern-Simons partition function that reveals the structure of poles in the Borel plane for $\Sigma(2,3,6n+5)$. In section \ref{sec:3}, we perform the resurgent analysis of Borel transformed SU(2) Chern-Simons partition function derived in the previous section. We specifically note the Stokes phenomenon and the different types of poles in the Borel plane where the non-perturbative contributions to the partition function are encoded. At last, We verify our analysis for $\Sigma(2,3,6n+5)$ by performing a resurgent analysis on $\Sigma(2,3,11)$ in full details in section \ref{sec:ex}.

\section{Exact partition function for $\Sigma(2,3,6n+5)$}
\label{sec:2}
The Brieskorn sphere $\Sigma(2,3,6n+5)$, an integer homology 3-sphere, is defined as
\begin{equation}
    \Sigma(2,3,6n+5)=\{(x,y,z)\in\mathbb{C}^3|x^2+y^3+z^{6n+5}=0\}\cap S^5
\end{equation}
where $S^5$ is a sphere in $\mathbb{C}^5$, $n\geq 1$, and $6n+5$ is a prime. (Note that all choices of $6n+5$ will maintain the pair-wise coprime relationship among $p_1,p_2,p_3$ in the Brieskorn sphere $\Sigma(p_1,p_2,p_3)$.)

We are interested in generating the homological blocks, $\hat{Z}_b$, which is a q-series invariant, of a given 3-manifold. This structure is related to our familiar WRT invariants of 3-manifold in the following way:
\begin{equation}
\label{eq:defZhat}
    Z_{\text{CS}}(M_3)=\frac{q-1}{i\sqrt{2k}q^{1/2}}\tau_k\left(M_3\right)=\frac{\tau_{k}(M_3)}{\tau_{k}(S^2\times S^1)}
\end{equation}
with the following normalization
\begin{equation*}
    \tau_k(S^3)=1,\qquad \tau_k(S^2\times S^1)=\sqrt{\frac{k}{2}}\frac{1}{\sin(\pi/k)},
\end{equation*}
s.t. we recover the standard physical normalization described in the literature
\begin{equation*}
    Z_{\text{CS}}(S^2\times S^1)=1,\qquad Z_{\text{CS}}(S^3)=\frac{2}{k}\sin\frac{\pi}{k},
\end{equation*}
where $k$ is the level of the Chern-Simons theory and will be complexified in our analysis. With this in mind, for plumbed manifolds, we then have the following conjectured relationship \cite{GPPV}
\begin{equation}
    Z_{SU(2)_k}\left(M_3\right)=\left(i\sqrt{2k}\right)^{b_1(M_3)-1}\sum_{a,b\in \text{Tor }H_1(M_3;\mathbb{Z})/\mathbb{Z}_2}e^{2\pi i k lk(a,a)}S_{ab}\hat{Z}_b(q)|_{q\to e^{2\pi i/k}}
\end{equation}
where we identify 
\begin{equation*}
    \hat{Z}_b(q)\in 2^{-c}q^{\Delta_b}\mathbb{Z}[[q]]\qquad \Delta_b\in \mathbb{Q},\qquad c\in \mathbb{Z}_+
\end{equation*}
s.t. $\hat{Z}_b(q)$ is the q-series generalisation of the WRT invariants. (The above definitions are assumed to be done with the gauge group $SU(2)$ but can be generalized to any connected gauge group.) However, we can greatly simplify this process of computing our homological blocks via making the following consideration: the newly defined homological blocks are analytic continuations of the WRT invariants and has the following categorification:
\begin{equation}
    \hat{Z}_b(q,t)=\sum_{i,j}q^it^j\text{dim}\mathcal{H}^{i,j}(M_3;a).
\end{equation}
Furthermore, if we consider the calculation of these homological blocks on a plumbed manifold, we have the following explicit form \cite{GMP}
\begin{multline}
\label{eq:qseriesgenerator}
    \hat{Z}_b(q)=q^{-\frac{3L+\sum_va_v}{4}}\cdot \text{p.v.}\int_{|z_v|=1}\prod_{v\in\text{Vertices}}\frac{dz_v}{2\pi iz_v}(z_v-1/z_v)^{2-\text{deg}(v)}\cdot \Theta_b^{-M}(z)\\
    \equiv q^{-\frac{3L+\sum_v a_v}{4}}\cdot \text{p.v.} \int_{|z_v|=1}\prod_{v\in\text{Vertices}}\frac{dz_v}{2\pi iz_v}\left(z_v-1/z_v\right)^2\\
    \times \prod_{(v_j,v_k)\in\text{Edges}}\frac{1}{\left(z_{v_j}-1/z_{v_j}\right)\left(z_{v_k}-1/z_{v_k}\right)}\cdot \Theta_b^{-M}(z)
\end{multline}
where we make the following identification
\begin{align*}
    \frac{1}{2}\text{p.v.}\int_{|z|=1}\frac{dz}{2\pi iz}(\dots)&=\lim_{\epsilon\to 0^+}\frac{1}{4}\left(\int_{|z|=1+\epsilon}+\int_{|z|=1-\epsilon}\right)\frac{dz}{2\pi iz}(\dots)\\
    \Theta_b^{-M}(z)&=\sum_{l\in 2M\mathbb{Z}^L+b}q^{-(l,M^{-1}l)/4}\cdot\prod_{i=1}^Lx_i^{l_i}\\
    b\in (2\text{Coker}M+\delta)/\mathbb{Z}_2,&\quad \delta\in \mathbb{Z}^L/2\mathbb{Z}^L
\end{align*}
and $(\cdot,\cdot)$ is the standard pairing on $\mathbb{Z}^l$. (Note the $\mathbb_2$ symmetry of b.)

Next, we focus on the resurgence aspect of these $\hat{Z}_b(q)$ invariants of 3-manifolds with the help of mock theta-functions which, explicitly, are Eichler integrals of weight-$3/2$ vector valued modular forms \cite{LZ,GMP,Hikami}. These mock modular forms are the holomorphic components of the harmonic weak Maass form. 
%A standard shorthand notation for q-Pochhamer symbol is given as 
%\begin{equation*}
%    (x)_n\equiv (x;q)_n=(1-x)(1-xq)\dots (1-xq^{n-1}).
%\end{equation*}
Mock modular theta-functions have the following building blocks
\begin{equation}
    \tilde{\Psi}_p^{(a)}(q):=\sum_{n=0}^{\infty}\psi_{2p}^{(a)}(n)q^{\frac{n^2}{4p}}\quad \in q^{\frac{a^2}{4p}}\mathbb{Z}[[q]]
\end{equation}
where 
\begin{equation*}
    \psi_{2p}^{(a)}(n)=
    \begin{cases}
    \pm 1 & n\equiv \pm a\text{ mod }2p,\\
    0 & \text{otherwise.}
    \end{cases}
\end{equation*}
We adapt the following shorthand notation for expressing linear combinations of the mock modular theta-functions
\begin{equation*}
    \tilde{\Psi}_p^{n_a(a)+n_b(b)+\dots}(q):=n_a\tilde{\Psi}_p^{(a)}(q)+n_b\tilde{\Psi}_p^{(b)}(q)+\dots
\end{equation*}
Using Borel transformation, the following has been identified in \cite{GMP}:
\begin{equation}
\label{eq:borelMock}
    \tilde{\Psi}_p^{(a)}(q)=\frac{\sqrt{k}}{2}\left(\int_{ie^{+i\delta}\mathbb{R}_+}+\int_{ie^{-i\delta}\mathbb{R}_-}\right)\frac{\sinh\left((p-a)\sqrt{\frac{2\pi i}{p}}\xi\right)}{\sinh\left(p\sqrt{\frac{2\pi i}{p}\xi}\right)}e^{-k\xi} \frac{d\xi}{\sqrt{\pi\xi}}.
\end{equation}
We can recover $\tilde{\Psi}_p^{(a)}(q)$ by taking the average of the generalized Borel sums \cite{GMP}
\begin{equation}
\label{eq:AverageBorel}
    Z_{\text{CS}}(q)=\frac{1}{2}\left[S_{\frac{\pi}{2}-\delta}Z_{\text{pert}}(k)+S_{\frac{\pi}{2}+\delta}Z_{\text{pert}}(k)\right]
\end{equation}
where the analyticity of partition function is now observable from the hyperbolic functions. Additionally, we can construct the linking matrix defined as
\begin{equation}
    M_{v_1,v_2}=
    \begin{cases}
    1& v_1,v_2\text{ connected},\\
    a_v& v_1=v_2=v,\\
    0& \text{ otherwise}.
    \end{cases}
    \qquad v_i\in\text{ Vertices of }\Gamma\cong \{1,\dots,L\},
\end{equation}
to help us compute $\hat{Z}$ from the generating functions. Note that with this definition, the linking matrix M can be constructed directly from reading data off of the plumbing graph of a 3-manifold. An intuitive way of associating a (connected) plumbing graph to a 3-manifold can be done when one views the plumbing graph as a link of unknots where the number of components in the link is equivalent to the number of vertices in the plumbing graph. This is beautifully illustrated in Fig. 5 and Sec. 3.4 of \cite{GPPV}. With a link now corresponding to a given connected plumbing graph, we can then conjecture that the 3-manifold which can be obtained via Dehn surgery from the link is now associated to the plumbing graph, i.e. visually, the process becomes
\begin{equation}
\label{eq:plumbGraphConstruct}
    \Gamma \xrightarrow{\text{\# of vertices }=\text{ \# of components}} L(\Gamma)\xrightarrow{\text{Dehn surgery}}M_3(L(\Gamma))\equiv M_3(\Gamma)
\end{equation}
where $\Gamma$ denotes the plumbing graph, $L(\Gamma)$ denotes the corresponding link of unknots, and $M_3(\Gamma)$ denotes the associated 3-manifold.
Pertaining to our discussion on Brieskorn homology spheres, an efficient and rigorous algorithm for computing the negative definite plumbing graphs of Seifert fibered rational homology spheres was put forth and proven in \cite{NR,OS}\footnote{It is also noteworthy that such a procedure was made more precise and specific for the case of $\Sigma(2,3,6n+1)$ in Sec. 2 of \cite{Durusoy}.}. An interesting relationship between the cokernel of M and the first homology group of the given link is \cite{Kucharski}
\begin{equation}
H_1(\Gamma,\mathbb{Z})\cong\text{Coker}M=\mathbb{Z}^L/M\mathbb{Z}^L
\end{equation}
and the number of elements in each set is $\text{det }M$.

Noting the relationship between WRT invariants and our $\hat{Z}$-invariants of 3-manifolds introduced in Eq. \eqref{eq:defZhat}, we can use the results introduced in \cite{Hikami} where we have for a Brieskorn sphere $\Sigma(p_1,p_2,p_3)$, if $1/p_1+1/p_2+1/p_3<1$, then 
\begin{equation}
    e^{\frac{2\pi i}{N}\frac{\phi(p_1,p_2,p_3)}{4}-\frac{1}{2}}\left(e^{\frac{2\pi i}{N}}-1\right)\tau_N(\Sigma(p_1,p_2,p_3))=\frac{1}{2}\tilde{\Phi}^{(1,1,1)}_{p}(1/N)
\end{equation}
where $\tilde{\Phi}_p^{(l_1,l_2,l_3)}(1/N)$ is a modular form with weight $3/2$ identified in Eq. (3.4) of \cite{Hikami}. This compact definition allows us to further verify, via the definition of the WRT invariants, the generating function for $\hat{Z}$-invariants of the SU(2) Chern-Simons partition function on a plumbed 3-manifold. In this case, we can then proceed to systematically compute the $\hat{Z}$-invariants of our interested Brieskorn spheres given that $1/2+1/3+1/(6n+5)<1,$ $\forall n\in\mathbb{Z}_+$. Thus, we have
\begin{equation}
\label{eq:ZcsMock}
    Z_{\text{CS}}(\Sigma(2,3,6n+5);q)=\frac{1}{i2\sqrt{2k}q^{\phi(n)}}\tilde{\Psi}_{2\cdot{3}\cdot{(6n+5)}}^{n_1\lambda_1+n_2\lambda_2+n_3\lambda_3+n_4\lambda_4}(q)
\end{equation}
where 
\begin{equation}
    \phi(n)=3-\frac{2}{\sqrt{3}}-\frac{1}{36n+30}\cdot{}\left(-1+18\cdot\sum_{i=1}^{6n+4}\cot\left(\frac{i\pi}{6n+5}\right)\cot\left(\frac{6i\pi}{6n+5}\right)\right)
\end{equation}
and \cite{Chun}
\begin{equation}
    \tilde{\Psi}_{2\cdot{3}\cdot{(6n+5)}}^{n_1\lambda_1+n_2\lambda_2+n_3\lambda_3+n_4\lambda_4}(q)=-\frac{1}{2}\sum_{\epsilon_1\epsilon_2\epsilon_3=\pm 1}\epsilon_1\epsilon_2\epsilon_3\tilde{\Psi}^{p_1p_2p_3(1+\sum_j\epsilon_j/p_j)}_{p_1p_2p_3}(q)
\end{equation}
for a given Brieskorn sphere $\Sigma(p_1,p_2,p_3)$. In particular, for a given Brieskorn sphere $\Sigma(p_1,p_2,p_3)$, we have the number of independent Eichler integrals present in the modular form, i.e. irreducible $SL(2,\mathbb{C})$ flat connections of the Chern-Simons partition function, is \cite{Hikami}
\begin{equation}
\label{eq:amountPoles}
    D=D(p_1,p_2,p_3)=\frac{(p_1-1)(p_2-1)(p_3-1)}{4}.
\end{equation}
We further note that this will introduce D amounts of triples $(l_1,l_2,l_3)$ where we can compute the Chern-Simons invariant of the Brieskorn sphere \cite{Fintushel}
\begin{equation}
    \text{CS}(\alpha)=-\frac{p_1p_2p_3}{4}\left(1+\sum_{j=1}^3\frac{l_j}{p_j}\right)^2\text{ mod }1.
\end{equation}
for flat connections in the theory \footnote{Note, the D triples $(l_1,l_2,l_3)\in\mathbb{Z}^3$ each satisfy the condition that $0< l_j< p_j$ specified in \cite{Hikami}. Furthermore, once the following five relationships are satisfied:
\begin{equation}
\begin{aligned}[c]
    l_2\text{ mod }2&=l_3\text{ mod 2},\\ \sum_{j=2}^3\frac{l_j}{p_j}&<1,\\ \sum_{j=1}^3\frac{l_j}{p_j}&>1,\\
\end{aligned}
\qquad
\begin{aligned}[c]
    \frac{l_1}{p_1}-\frac{l_2}{p_2}+\frac{l_3}{p_3}&<1,\\ \frac{l_1}{p_1}+\frac{l_2}{p_2}-\frac{l_3}{p_3}&<1,
\end{aligned}
\end{equation}
then the integer triple $(l_1,l_2,l_3)$ has a one-to-one correspondence with elements in the representation space of conjugacy classes of irreps of $\pi_1(\Sigma(p_1,p_2,p_3))$, defined later explicitly in Eq. \eqref{eq:fundRep}, into SU(2) \cite{Fintushel,KirkKlassen}. Given that this representation class is finite, there exists a finite set of triples $(l_1,l_2,l_3)$ (D amount) which can be used to compute the Chern-Simons invariants for a given Brieskorn sphere.}. Concretely, for our set of Brieskorn spheres $\Sigma(2,3,6n+5)$, we have
\begin{equation}
\label{eq:lambdas}
    \lambda_1= 6n-1,\quad 
    \lambda_2= 6n+11,\quad 
    \lambda_3= 30n+19,\quad
    \lambda_4= 30n+31,
\end{equation}
and the corresponding coefficients are
\begin{equation}
\label{eq:ns}
    n_1= 1,\quad n_2= -1,\quad n_3=-1,\quad n_4= 1.
\end{equation}
This can be verified as we have that all Brieskorn spheres are integral homology spheres which indicates $H_1(\Sigma(2,3,6n+5))=0$ and the typically represented $\hat{Z}_b(q)$ is expressed for $\Sigma(2,3,6n+5)$ with specifically $b=0$.

Using the Borel transformation results obtained for the mock modular theta-functions in Eq. \eqref{eq:borelMock}, we can obtain the resurgence form of the $\hat{Z}$-invariants shown in Eq. \eqref{eq:ZcsMock}:
\begin{multline}
\label{eq:resurgence1}
    Z_{\text{CS}}(\Sigma(2,3,6n+5))\sim \frac{\sinh((30n+31)z)}{\sinh((36n+30)z)}
    -\frac{\sinh((30n+19)z)}{\sinh((36n+30)z)}
    \\-\frac{\sinh((6n+11)z)}{\sinh((36n+30)z)}+\frac{\sinh((6n-1)z)}{\sinh((36n+30)z)}\\
    =\frac{4\sinh(6z)\sinh((12n+10)z)\sinh((18n+15)z)}{\sinh((36n+30)z)},
\end{multline}
where we identify 
\begin{equation*}
    Z_{\text{CS}}^{\text{resurgence}}(\Sigma(2,3,6n+5);z)\equiv \frac{4\sinh(6z)\sinh((12n+10)z)\sinh((18n+15)z)}{\sinh((36n+30)z)}.
\end{equation*} 
Note, the results shown above is also a special case of Eq. (2.101) in \cite{Chung2018} where $M_3=\Sigma(2,3,6n+5)$. We notice that the poles in the R.H.S. expression of Eq. \eqref{eq:resurgence1} will eventually give rise to the non-perturbative contributions to the Chern-Simons partition function on the manifold. In particular, the residues of these poles match up with the associated Chern-Simons instanton actions. Carefully computing the residues corresponding to each pole will allow one to distinguish the properties of the pole, i.e. whether this contributes to an abelian flat connection, non-abelian flat connection, or complex flat connection. Note that complex flat connections has a complex classical Chern-Simons action where this manifests itself as a pole in the Borel transformation of the perturbative expansion around the complex flat connection. These complex flat connections then do not contribute to the Chern-Simons partition function anymore.

\section{Resurgent analysis for $\Sigma(2,3,6n+5)$}
\label{sec:3}
Here, we will apply our previous arguments onto our interested family of Brieskorn sphere $\Sigma(2,3,6n+5)$ where $n\in\mathbb{Z}_+$ and $6n+5$ is a prime. Given that the general exact Chern-Simons partition function of these Brieskorn spheres can still be expressed in terms of mock modular theta-functions as found in Eq. \eqref{eq:ZcsMock}, we follow the similar resurgent analysis structure as done before in the literature \cite{GMP,Chun}. Hence, we start with analyzing the structures of poles exhibited in $Z_{\text{CS}}^{\text{resurgence}}(\Sigma(2,3,6n+5);z)$ found in Eq. \eqref{eq:resurgence1}. Notice that $Z_{\text{CS}}^{\text{resurgence}}(\Sigma(2,3,6n+5);z)$ has simple poles at $z=m\pi i/(36n+30),$ $\forall m\in{\mathbb{Z}}$ and is non-divisible by 2,3,$6n+5$. We then proceed to compute the contributions to the averaged Borel sums from the poles which are aligned along the imaginary axis in the Borel plane. 

The classical solution for Chern-Simons action defined in Eq. \eqref{eq:chernsimonsFunc} must be flat connections according to the EOMs. Furthermore, our partition function necessarily sums over the moduli space of flat connections on $M_3$ of which are the saddle points of the Chern-Simons action. For our particular case, since we are dealing with a Brieskorn homology sphere, the moduli space of flat connections, $\mathcal{M}_{flat}\left(\Sigma(2,3,6n+5),\text{SL}(2,\mathbb{C})\right)$, suggest there to exist $3n+2$ irreducible $\text{SL}(2,\mathbb{C})$ flat connections along with an abelian trivial flat connection which we will call $\alpha_0$ which is in agreement with Eq. \eqref{eq:amountPoles} found from \cite{Hikami} and can be verified numerically in Table \ref{table:Poles}.
\begin{table}[h]
\centering
\begin{tabular}{ |p{2.0cm}||p{11.0cm}|  }
%\centering
\hline 
Brieskorn Sphere & Poles\\
 \hline
 $\Sigma(2,3,11)$ & $1,5,7,13,19$\\
 \hline
 $\Sigma(2,3,17)$ & $1,5,7,11,13,19,25,31$\\
 \hline
 $\Sigma(2,3,23)$ & $1,5,7,11,13,17,19,25,31,37,43$\\
 \hline
 $\Sigma(2,3,29)$ & $1,5,7,11,13,17,19,23,25,31,37,43,49,55$\\
 \hline
 $\Sigma(2,3,41)$ & $1,5,7,11,13,17,19,23,25,29,31,35,37,43,49,55,61,67,73,79$\\
 \hline
\end{tabular}
\caption{The first five of Brieskorn spheres $\Sigma(2,3,6n+5)$ and their corresponding family of poles that contribute to their irreducible $SL(2,\mathbb{C})$ connections.}
\label{table:Poles}
\end{table}
\begin{table}[h]
\centering
\begin{tabular}{ |p{2.0cm}||p{8.0cm}|p{4cm}|  }
\hline 
Brieskorn Sphere & Flat & Complex\\
 \hline
 $\Sigma(2,3,11)$ & $1,7,13,19$ & $5$\\
 \hline
 $\Sigma(2,3,17)$ & $1,7,13,19,25,31$ & $5,11$\\
 \hline
 $\Sigma(2,3,23)$ & $1,7,13,19,25,31,37,43$ & $5,11,17$\\
 \hline
 $\Sigma(2,3,29)$ & $1,7,13,19,25,31,37,43,49,55$ & $5,11,17,23$\\
 \hline
 $\Sigma(2,3,41)$ & $1,7,13,19,25,31,37,43,49,55,61,67,73,79$ & $5,11,17,23,29,35$\\
 \hline
 \end{tabular}
\caption{The first five of Brieskorn spheres $\Sigma(2,3,6n+5)$ and their poles corresponding to flat and complex $SL(2,\mathbb{C})$ connections.}
\label{table:PolesLab}
\end{table}

Furthermore, for an irreducible flat connection $\alpha_{\text{flat}}$, the Chern-Simons instanton action is
\begin{equation}
    \text{CS}(\alpha_{\text{flat}})=-\frac{18n+15}{2}\left(1+\sum_i \frac{2x_1+3x_2+(6n+5)x_3}{36n+30}\right)
\end{equation}
where the triples $(x_1,x_2,x_3)$ satisfy the relations given by the general fundamental group for Brieskorn spheres \cite{Hikami}:
\begin{equation}
\label{eq:fundRep}
    \pi_1\left(\Sigma(p_1,p_2,p_3)\right)=\langle x_1,x_2,x_3,h\rvert h\text{ central},x_k^{p_k}=h^{-q_k}\text{ for }k=1,2,3,x_1x_2x_3=1\rangle
\end{equation}
where $q_k\in \mathbb{Z}$ s.t.
\begin{equation}
    P\left(-1+\sum_{k=1}^3\frac{q_k}{p_k}\right)=0\Rightarrow P\sum_{k=1}^3\frac{q_k}{p_k}=1
\end{equation}
and $P=P(p_1,p_2,p_3)=p_1p_2p_3$. Thus, concretely, for a general Brieskorn sphere $\Sigma(2,3,6n+5)$, in addition to Chern-Simons functional associated to the abelian trivial flat connection, $\alpha_0$, the full set of classical Chern-Simons functionals are those s.t.
\begin{dmath}
    \text{CS}(\alpha)\in \mathcal{M}_{\text{non-abelian flat}}\cup \mathcal{M}_{\text{complex}}\cup \{\text{CS}(\alpha_0)\}
\end{dmath}
where
\begin{dmath}
\label{eq:nonabelianflat}
    {\mathcal{M}_{\text{non-abelian flat}}:=}\left\{-\frac{\left[10(3\beta-1)+5\right]^2}{72n+60},-\frac{\left[10(3\beta-2)+3\right]^2}{72n+60},
    -\frac{\left[10(3\beta-2)+9\right]^2}{72n+60},-\frac{\left[10(3\beta-3)+1\right]^2}{72n+60},-\frac{\left[10(3\beta-3)+7\right]^2}{72n+60}\ \bigg\rvert\ {0<\beta<\gamma,\beta\in\mathbb{Z}}\right\}\bigcup\left\{-\frac{[10\gamma+\kappa_i]^2}{72n+60}\ \bigg\rvert\ {\kappa\in \{1,7,3,9\}}\right\}
\end{dmath}
of which $\gamma=\floor*{\frac{3n+2}{5}},\ i\in\{1,...,(3n+2)\text{ mod }5\}$ and
\begin{equation}
\label{eq:complex}
    \mathcal{M}_{\text{complex}}:=\left\{-\frac{(6k+5)^2}{72n+60}\ \bigg\rvert\  k\in[0,n-1]\right\}.
\end{equation}
We can similarly verify these results numerically and these are presented in Table \ref{table:PolesLab}. (Note, the purpose of the second set in $\mathcal{M}_{\text{non-abelian flat}}$ is to simply choose from the (ordered) set $\{1,7,3,9\}$ given $(3n+2)\text{ mod }5$.)

To further our argument, we need to complete our description of the members that belong to each pole. In particular, we discover that to each irreducible $SL(2,\mathbb{C})$ connection, there are always 8 poles associated under the same connection with the same behavior and overall factor. The basic algorithm of computing the family of poles that are associated to the Chern-Simons instanton action 
\begin{equation*}
    \text{CS}(\alpha)=-\frac{n^2}{72n+60},
\end{equation*}
is to then compute all the poles, $\alpha'$, that satisfy:
\begin{equation}
    \text{CS}(\alpha'):=\frac{x^2\text{ mod }(72n+60)}{72n+60}=-\text{CS}(\alpha),\quad \forall x\in{[0,36n+30]}.
\end{equation}

Now, using our insight gained from Eq. \eqref{eq:fundRep}, we can categorize these poles into flat and complex irreducible $SL(2,\mathbb{C})$ connections. For a given Brieskorn sphere, we have $2(n+1)$ flat irreducible connections and $n$ complex irreducible connections as verified in Table \ref{table:PolesLab}.
\begin{table*}[h]
\centering
    \begin{subtable}[ht]{\textwidth}
       \caption{$\Sigma(2,3,17)$}
       \label{tab:Sig2317}
        \centering
        \begin{tabular}[t]{l |l |l }
        Poles & Residues & Overall Factor\\
        \hline \hline
        $\{1,35,67,101,103,137,169,203\}$ & $\{-1,-1,-1,-1,+1,+1,+1,+1\}$& $\frac{i}{51}[\sin(11\pi/102)-\sin(23\pi/102)]$\\
        $\{5,29,73,97,107,131,175,199\}$ & $\{-1,+1,+1,-1,+1,-1,-1,+1\}$& $\frac{i}{51}[\sin(2\pi/51)+\sin(13\pi/102)]$\\
        $\{7,41,61,95,109,143,163,197\}$ & $\{-1,-1,-1,-1,+1,+1,+1,+1\}$& $\frac{i}{51}[\cos(4\pi/51)+\sin(25\pi/102)]$\\
        $\{11,23,79,91,113,125,181,193\}$ & $\{+1,-1,-1,+1,-1,+1,+1,-1\}$& $\frac{i}{51}[\cos(\pi/51)+\sin(19\pi/102)]$\\
        $\{13,47,55,89,115,149,157,191\}$ & $\{+1,+1,+1,+1,-1,-1,-1,-1\}$& $\frac{i}{51}[\cos(5\pi/51)+\sin(7\pi/102)]$\\
        $\{19,49,53,83,121,151,155,185\}$ & $\{+1,+1,+1,+1,-1,-1,-1,-1\}$& $\frac{i}{51}[\cos(11\pi/51)-\sin(5\pi/102)]$\\
        $\{25,43,59,77,127,145,161,179\}$ & $\{-1,-1,-1,-1,+1,+1,+1,+1\}$& $\frac{i}{51}[\cos(7\pi/51)+\cos(10\pi/51)]$\\
        $\{31,37,65,71,133,139,167,173\}$ & $\{+1,+1,+1,+1,-1,-1,-1,-1\}$& $\frac{i}{51}[\cos(8\pi/51)+\sin(\pi/102)]$
       \end{tabular}
    \end{subtable}
    \begin{subtable}[ht]{\textwidth}
        \caption{$\Sigma(2,3,23)$}
        \label{tab:Sig2323}
        \centering
        \begin{tabular}[t]{l |l |l }
        Poles & Residues & Overall Factor\\
        \hline \hline
        $\{1,47,91,137,139,185,229,275\}$ & $\{-1,-1,-1,-1,+1,+1,+1,+1\}$& $\frac{i}{69}[\sin(17\pi/138)-\sin(29\pi/138)]$\\
        $\{5,41,97,133,143,179,235,271\}$ & $\{-1,+1,+1,-1,+1,-1,-1,+1\}$& $\frac{i}{69}[\cos(8\pi/69)+\sin(7\pi/138)]$\\
        $\{7,53,85,131,145,191,223,269\}$ & $\{-1,-1,-1,-1,+1,+1,+1,+1\}$& $\frac{i}{69}[\cos(2\pi/69)+\sin(19\pi/138)]$\\
        $\{11,35,103,127,149,173,241,265\}$ & $\{+1,-1,-1,+1,-1,+1,+1,-1\}$ & $\frac{i}{69}[\cos(10\pi/69)+\cos(13\pi/69)]$\\
        $\{13,59,79,125,151,197,217,263\}$ & $\{+1,+1,+1,+1,-1,-1,-1,-1\}$ & $\frac{i}{69}[\cos(7\pi/69)+\cos(16\pi/69)]$\\
        $\{17,29,109,121,155,167,247,259\}$ & $\{-1,+1,+1,-1,+1,-1,-1,+1\}$ & $\frac{i}{69}[\cos(5\pi/69)+\sin(13\pi/138)]$\\
        $\{19,65,73,119,157,203,211,257\}$ & $\{-1,-1,-1,-1,+1,+1,+1,+1\}$ & $\frac{i}{69}[\cos(11\pi/69)+\sin(\pi/138)]$\\
        $\{25,67,71,113,163,205,209,251\}$ & $\{-1,-1,-1,-1,+1,+1,+1,+1\}$ & $\frac{i}{69}[\cos(17\pi/69)-\sin(11\pi/138)]$\\
        $\{31,61,77,107,169,199,215,245\}$ & $\{+1,+1,+1,+1,-1,-1,-1,-1\}$ & $\frac{i}{69}[\cos(\pi/69)+\sin(25\pi/138)]$\\
        $\{37,55,83,101,175,193,221,239\}$ & $\{-1,-1,-1,-1,+1,+1,+1,+1\}$ & $\frac{i}{69}[\cos(4\pi/69)+\sin(31\pi/138)]$\\
        $\{43,49,89,95,181,187,227,233\}$ & $\{+1,+1,+1,+1,-1,-1,-1,-1\}$ & $\frac{i}{69}[\cos(14\pi/69)-\sin(5\pi/138)]$
        \end{tabular}
     \end{subtable}
     \caption{The residues and overall factors of the corresponding family of poles for $\Sigma(2,3,17)$ and $\Sigma(2,3,23)$.}
     \label{tab:poles}
\end{table*}
We notice that the complex connections for the Brieskorn homology sphere $\Sigma(2,3,6n+5)$ is associated to the Chern-Simons instanton action
\begin{equation}
    \text{CS}(\alpha)_{\text{complex}}=-\frac{(6k+5)^2}{72n+60},\quad k\in[0,n-1]
\end{equation}
which confirms our set of complex connections defined earlier in Eq. \eqref{eq:complex}. The next immediate step is to compute the overall factor of these classical Chern-Simons functionals. For a given general Brieskorn sphere of the structure $\Sigma(2,3,6n+5)$, we have the Chern-Simons functional's overall factor is
\begin{equation}
\label{eq:tableMotiv}
    \text{CS}(\alpha)=-\frac{k^2}{72n+60}\quad \xrightarrow[]{\text{Overall Factor}}\quad \text{Res}\left(k,Z_{\text{CS}}^{\text{resurgence}}(\Sigma(2,3,6n+5))\right).
\end{equation}
This can be evaluated explicitly as
\begin{equation*}
\text{Res}\left(k,Z_{\text{CS}}^{\text{resurgence}}(\Sigma(2,3,6n+5))\right)
    =\frac{i}{18n+15}\cdot{}
    \begin{cases}
       \sin\left(\frac{a\pi}{36n+30}\right)-\sin\left(\frac{b\pi}{36n+30}\right)\\
       \cos\left(\frac{a\pi}{36n+30}\right)+\cos\left(\frac{b\pi}{36n+30}\right)\\
       \cos\left(\frac{a\pi}{36n+30}\right)+\sin\left(\frac{b\pi}{36n+30}\right)\\
       \cos\left(\frac{a\pi}{36n+30}\right)-\sin\left(\frac{b\pi}{36n+30}\right)\\
    \end{cases}
\end{equation*}
where $a,b\in\mathbb{Z}$ and $(a+b) \text{ mod } k=0$ or $(a-b) \text{ mod } k=0$ which vary depends on the choice of n.

Notice that each of these instanton values originate from the poles along the imaginary axis. For every pole, it contributes to the overall Chern-Simons partition function by a factor of $e^{2\pi i \text{CS}(\alpha_i)}$ where $\text{CS}(\alpha_i)$ corresponds to the Chern-Simons instanton value of the given pole, i.e. Eq.\eqref{eq:decompCSfull}. Therefore, we have a complete description of $Z_{\text{CS}}$ in terms of the non-abelian flat connections on $\Sigma(2,3,6n+5)$. Furthermore, we have
\begin{equation}
    n_{\alpha,0}=
    \begin{cases}
    1,&\alpha=(\alpha_0,0)\\
    \frac{1}{2}m_{\alpha}^{\alpha_0,0},&\text{otherwise}
    \end{cases}
\end{equation}
where we have
\begin{equation}
    m_{\beta}^{\alpha_0,0}=
    \begin{cases}
    +1, & \beta=\left(\alpha\in{\mathcal{M}_{non-abelian flat}(G_{\mathbb{C}},\Sigma(2,3,6n+5))},-\frac{-n^2}{72n+60}\right), n=n_i\text{ mod }72n+60\\
    -1, & \beta=\left(\alpha\in{\mathcal{M}_{non-abelian flat}(G_{\mathbb{C}},\Sigma(2,3,6n+5))},-\frac{-n^2}{72n+60}\right),n=n_j\text{ mod }72n+60\\
    0,&\text{otherwise.}
    \end{cases}
\end{equation}
where $j=|9-i|$ and $n_a\in\{\tilde{n}_a\ \rvert\ \tilde{n}_a\text{ mod }72n+60=n\text{ mod }72n+60\}$. The specific pattern is dictated by the residue value of each pole on different Brieskorn spheres. Specific examples are given in Table \ref{tab:poles}.
Therefore, we can now account for the contributions coming from each pole. For instance, poles that have the same classical Chern-Simons instanton action $\text{CS}(\alpha)=-k^2/(72n+60)$ contribute to $Z_{\text{CS}}$ as
\begin{dmath}
\label{eq:resContCS}
    \text{Res}(k,Z_{\text{CS}}^{\text{resurgence}}(\Sigma(2,3,6n+5)))\cdot{}\left(\sum_{\substack{m\text{ mod }(72n+60)=k\\n=\pm k\text{ mod (36n+30)}}}\pm 1\right)e^{2\pi ik\text{CS}(\alpha)}=\text{Res}(k,Z_{\text{CS}}^{\text{resurgence}}(\Sigma(2,3,6n+5)))\cdot{}\left(\sum_{m\text{ mod }(72n+60)=k}1-\frac{m}{18n+15}\right)e^{2\pi ik\text{CS}(\alpha)}
\end{dmath}
where we obtain the R.H.S. via the zeta-function regularization: $\sum_{n=\pm a \text{ mod }2p}\pm 1=1-a/p$.
In particular, for $\Sigma(2,3,6n+5)$, we get
\begin{equation}
    \text{Eq. }\eqref{eq:resContCS}=2\text{Res}(k,Z_{\text{CS}}^{\text{resurgence}}(2,3,6n+5))e^{2\pi ik \text{CS}(\alpha)}
\end{equation}
Thus, we arrive at our main conclusion of this paper:
\begin{equation}
\label{eq:result}
    Z_{\text{CS}}(\Sigma(2,3,6n+5))=\sum_{\alpha\in \mathcal{M}_{\text{nonabelian flat}}(G_{\mathbb{C}},\Sigma(2,3,6n+5))}2\text{Res}(k,Z_{\text{CS}}^{\text{resurgence}}(2,3,6n+5))e^{-2\pi ik\text{CS}(\alpha)}
\end{equation}
where we define k as the usual poles corresponding to the irreducible flat connection $\alpha$: $\text{CS}(\alpha)=-k^2/(72n+60)$ and the set $\mathcal{M}_{\text{nonabelian flat}}(G_{\mathbb{C}},\Sigma(2,3,6n+5))$ as in Eq. \eqref{eq:nonabelianflat}. We can verify our conclusion Eq. \eqref{eq:result} via the following example $\Sigma(2,3,11)$ where $n=1$.

\section{Example: $\Sigma(2,3,11)$}
\label{sec:ex}
The Brieskorn homology sphere $\Sigma(2,3,11)$ is defined as
\begin{equation}
    \Sigma(2,3,11)=\{(x,y,z)\in\mathbb{C}^3|x^{2}+y^{3}+z^{11}=0\}\cap S^5.
\end{equation}
We can construct the following negative definite plumbing graph $\Gamma$ of our Brieskorn sphere $\Sigma(2,3,11)$ via Eq. \eqref{eq:plumbGraphConstruct} or similarly via the algorithm in \cite{NR}.
\begin{figure}[htpb]
\begin{subfigure}[htpb]{\textwidth}
  \centering
  \begin{tikzpicture}[>=stealth',join=bevel,font=\sffamily,auto,on grid,decoration={markings,
    mark=at position .5 with \arrow{>}}]
    \coordinate (Lnode) at (180:2cm);
    \coordinate (Unode) at (90:1cm);
    \coordinate (Rnode) at (0:6cm);
    \coordinate (originNode) at (0:0cm);

    \draw[-, very thick] (Lnode.south) -- (0,0) node[below,pos=0]{-2} node[below,pos=0.5]{-2};

    \draw[-, very thick] (Unode.south) -- (0,0) node[pos=0]{-2};

    \draw[-, very thick] (Rnode.south) -- (0,0) node[pos=0]{-3} node[pos=0.2]{-2} node[pos=0.4]{-2} node[pos=0.6]{-2} node[pos=0.8]{-2} node[pos=1]{-2};

    \draw[fill] (barycentric cs:Lnode=1.0,originNode=0) circle (2pt);
    \draw[fill] (barycentric cs:Lnode=0.5,originNode=0.5) circle (2pt);

    \draw[fill] (barycentric cs:Unode=1.0,originNode=0) circle (2pt);

    \draw[fill] (barycentric cs:Rnode=1.0,originNode=0) circle (2pt);
    \draw[fill] (barycentric cs:Rnode=0.8,originNode=0.2) circle (2pt);
    \draw[fill] (barycentric cs:Rnode=0.6,originNode=0.4) circle (2pt);
    \draw[fill] (barycentric cs:Rnode=0.4,originNode=0.6) circle (2pt);
    \draw[fill] (barycentric cs:Rnode=0.2,originNode=0.8) circle (2pt);
    \draw[fill] (barycentric cs:Rnode=0,originNode=1) circle (2pt);

  \end{tikzpicture}
  \caption{Brieskorn sphere $\Sigma(2,3,11)$} 

\end{subfigure}
\begin{subfigure}[htpb]{\textwidth}
  \centering
  \begin{tikzpicture}[>=stealth',join=bevel,font=\sffamily,auto,on grid,decoration={markings,
    mark=at position .5 with \arrow{>}}]
    \coordinate (Lnode) at (180:2cm);
    \coordinate (Unode) at (90:1cm);
    \coordinate (Rnode) at (0:6cm);
    \coordinate (originNode) at (0:0cm);

    \draw[-, very thick] (Lnode.south) -- (0,0) node[below,pos=0]{(3)} node[below,pos=0.5]{(2)};

    \draw[-, very thick] (Unode.south) -- (0,0) node[pos=0]{(4)};

    \draw[-, very thick] (Rnode.south) -- (0,0) node[pos=0]{(9)} node[pos=0.2]{(8)} node[pos=0.4]{(7)} node[pos=0.6]{(6)} node[pos=0.8]{(5)} node[pos=1]{(1)};

    \draw[fill] (barycentric cs:Lnode=1.0,originNode=0) circle (2pt);
    \draw[fill] (barycentric cs:Lnode=0.5,originNode=0.5) circle (2pt);

    \draw[fill] (barycentric cs:Unode=1.0,originNode=0) circle (2pt);

    \draw[fill] (barycentric cs:Rnode=1.0,originNode=0) circle (2pt);
    \draw[fill] (barycentric cs:Rnode=0.8,originNode=0.2) circle (2pt);
    \draw[fill] (barycentric cs:Rnode=0.6,originNode=0.4) circle (2pt);
    \draw[fill] (barycentric cs:Rnode=0.4,originNode=0.6) circle (2pt);
    \draw[fill] (barycentric cs:Rnode=0.2,originNode=0.8) circle (2pt);
    \draw[fill] (barycentric cs:Rnode=0,originNode=1) circle (2pt);

  \end{tikzpicture}
  \caption{Labeled Brieskorn sphere $\Sigma(2,3,11)$} 

\end{subfigure}
\caption{(a) The negative definite plumbing graph for the Brieskorn sphere $\Sigma(2,3,11)$ and (b) the labeling of the vertices for creating the linking matrix. (Note, while the labeling scheme is arbitrary and may lead to different explicit constructions of the linking matrix, the computed $\hat{Z}$ from Eq. \eqref{eq:qseriesgenerator} will nevertheless be the same.)}
\label{fig:Brieskorn2311}
\end{figure}
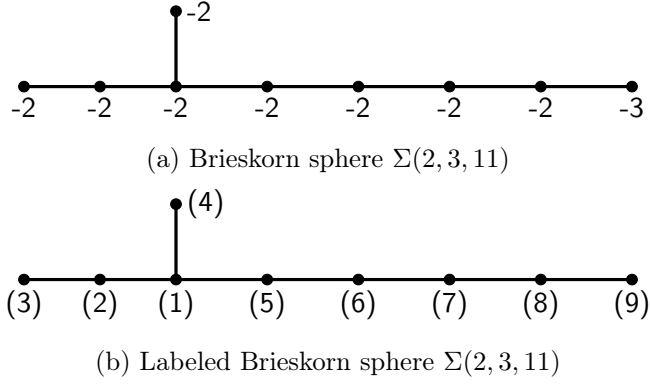
With information from Fig. \ref{fig:Brieskorn2311}, we can then obtain the linking matrix for $\Sigma(2,3,11)$ as:
\begin{equation}
M=
    \begin{pmatrix}
    -2&1&0&1&1&0&0&0&0\\
    1&-2&1&0&0&0&0&0&0\\
    0&1&-2&0&0&0&0&0&0\\
    1&0&0&-2&0&0&0&0&0\\
    1&0&0&0&-2&1&0&0&0\\
    0&0&0&0&1&-2&1&0&0\\
    0&0&0&0&0&1&-2&1&0\\
    0&0&0&0&0&0&1&-2&1\\
    0&0&0&0&0&0&0&1&-3
    \end{pmatrix}
\end{equation}
where, in particular, we have $L=9$, Vertices $v_i\in\{v_1,v_2,\dots,v_9\}$, and edges $(1,2),(2,3),$ $(1,4),(1,5),(5,6),(6,7),(7,8),(8,9)$ along with $\det M=-1$ and $\text{Coker}M=\{0\}$ which remains valid for all types of Brieskorn homology spheres. Now, we first conjecture the following from the ingredients gained from the linking matrix and the generating function Eq. \eqref{eq:qseriesgenerator}
\begin{dmath}
    \hat{Z}(\Sigma(2,3,11){)\sim  \tilde{\Psi}_{66}^{n_1(5)+n_2(17)+n_3(49)+n_4(61)}(q)}
    =\sum_{n=0}^{\infty}n_1\psi_{132}^{(5)}(n)q^{n^2/264}+n_2\psi_{132}^{(17)}(n)q^{n^2/264}+n_3\psi_{132}^{(49)}(n)+n_4\psi_{132}^{(61)}(n)q^{n^2/132}\\
    =n_1\left(q^{5^2/264}-q^{127^2/264}+q^{137^2/264}+\dots\right)+n_2\left(q^{17^2/264}-q^{115^2/264}+q^{149^2/264}+\dots\right)\\
    +n_3\left(q^{49^2/264}-q^{83^2/264}+q^{181^2/264}\right)+n_4\left(q^{61^2/264}-q^{71^2/264}+q^{193^2/264}\right)\\
    =q^{25/264}\left(n_1+n_2q+n_3q^9+n_4q^{14}-n_4q^{19}-n_3q^{26}+\dots\right).
\end{dmath}
This will give us the obvious verification of q-series for the Brieskorn homology sphere
\begin{equation*}
    \hat{Z}(\Sigma(2,3,11))\in q^{\Delta_b}\mathbb{Z}[[q]],\qquad \text{ with } n_1,n_2,n_3,n_4\in\mathbb{Z}.
\end{equation*}
Now, via the usual definition of WRT invariants using $\Psi_{p_1p_2p_3}^{(1,1,1)}(q)$ for general Brieskorn sphere $\Sigma(p_1,p_2,p_3)$ with $1/p_1+1/p_2+1/p_3<1$ applied onto $\Sigma(2,3,6n+5)$ conjectured in Eq. \eqref{eq:ZcsMock}, we identify
\begin{equation*}
    n_1= +1, \qquad n_2= -1, \qquad n_3= -1, \qquad n_4= +1.
\end{equation*}
Thus, we have, up to normalization, 
\begin{equation}
    \hat{Z}(\Sigma(2,3,11))\cong q^{25/264}\left(1-q-q^9+q^{14}-q^{19}+q^{26}+\dots\right)
\end{equation}
which is in agreement with Eq. \eqref{eq:lambdas} and Eq. \eqref{eq:ns}.
With normalization, we can write the exact Chern-Simons partition function as
\begin{align}
\begin{split}
    Z_{CS}(\Sigma(2,3,11))&=\frac{1}{i2\sqrt{2k}q^{421/264}}q^{25/264}\left(1-q-q^9+q^{14}-q^{19}+q^{26}+\dots\right)\\
    &=\frac{1}{i2\sqrt{2k}}q^{-3/2}\left(1-q-q^9+q^{14}-q^{19}+q^{26}+\dots\right).
\end{split}
\end{align}

Performing the resurgent analysis on $Z_{\text{CS}}(\Sigma(2,3,11))$ we obtain
\begin{equation}
\label{eq:resurgenta}
    Z_{\text{CS}}(\Sigma(2,3,11))\sim \frac{\sinh(61z)}{\sinh(66z)}-\frac{\sinh(17z)}{\sinh(66z)}-\frac{\sinh(49z)}{\sinh(66z)}+\frac{\sinh(5z)}{\sinh(66z)}.
    %\sim\frac{1}{\sinh(66z)}.
\end{equation}
We take note that the R.H.S. of Eq. \eqref{eq:resurgenta} has simple poles at $z=n\pi i/66$ for $n\in \mathbb{Z}$ and non-divisible by $2,3,11$. Then from the averaged Borel sum, we get that the poles are aligned on the imaginary axis and we can choose the same integration contours as in Eq. \eqref{eq:AverageBorel}.

The classical Chern-Simons functionals (classical instanton actions) take the following values (mod 1):
\begin{align}
\begin{split}
    \text{CS}(\alpha_0)=0,\quad \text{CS}(\alpha_1)&=-\frac{1}{264},\quad
    \text{CS}(\alpha_2)=-\frac{5^2}{264},\\
    \text{CS}(\alpha_3)=-\frac{7^2}{264},\quad
    \text{CS}(\alpha_4)&=-\frac{13^2}{264},\quad
    \text{CS}(\alpha_5)=-\frac{19^2}{264}.
\end{split}
\end{align}
where $\alpha_0$ is the trivial (abelian) flat connection. We have for our particular Brieskorn homology sphere, $\Sigma(2,3,11)$, we have a total of $(2-1)\cdot(3-1)\cdot(11-1)/4=5$ irreducible SL$(2,\mathbb{C})$ connections which agrees with Eq. \eqref{eq:amountPoles}. These can be found by considering the poles contribution to the Borel sums with residues accounted by the Stokes phenomenon. Hence, there are 5 groups of poles at $z=\frac{n\pi i}{p}$, with n counted modulo $2p=132$
\begin{itemize}
    \item $n=1,23,43,65,67,89,109,131$ have $\text{CS}=-\frac{1^2}{264}$ and residues $\{+1,+1,+1,+1,-1,-1,-1,-1\}$ respectively, up to an overall factor of $\frac{i}{33}\left[\cos\left(\frac{8\pi}{33}\right)-\sin\left(\frac{5\pi}{66}\right)\right]$;
    \item $n=5,17,49,61,71,83,115,127$ have $\text{CS}=-\frac{5^2}{264}$ and residues $\{-1,+1,+1,-1,+1,-1,-1,+1\}$ respectively, up to an overall factor $\frac{i}{33}\left[\cos\left(\frac{4\pi}{33}\right)+\sin\left(\frac{7\pi}{33}\right)\right]$;
    \item $n=7,29,37,59,73,95,103,125$ have $\text{CS}=-\frac{7^2}{264}$ and residues $\{-1,-1,-1,-1,+1,+1,+1,+1\}$ respectively, up to an overall factor $\frac{i}{33}\left[\cos\left(\frac{\pi}{33}\right)+\sin\left(\frac{13\pi}{66}\right)\right]$;
    \item $n=13,31,35,53,79,97,101,119$ have $\text{CS}=-\frac{13^2}{264}$ and residues $\{-1,-1,-1,-1,+1,+1,+1,+1\}$ respectively, up to an overall factor $\frac{i}{33}\left[\cos\left(\frac{5\pi}{33}\right)+\sin\left(\frac{\pi}{66}\right)\right]$;
    \item $n=19,25,41,47,85,91,107,113$ have $\text{CS}=-\frac{19^2}{264}$ and residues $\{+1,+1,+1,+1,-1,-1,-1,-1\}$ respectively, up to an overall factor $\frac{i}{33}\left[\cos\left(\frac{2\pi}{33}\right)+\sin\left(\frac{7\pi}{66}\right)\right]$.
\end{itemize}
From these poles analysis, we realize that $\alpha_1,\alpha_3,\alpha_4,\alpha_5$ are the four non-abelian flat connections that can be conjugated inside SU(2). In other words, these are the four irreducible flat connections in SU(2) of which $n_{\alpha}=1$ in the perturbative expansion
\begin{equation}
    Z_{\text{CS}}(\Sigma(2,3,11))=\sum_{\alpha\in\mathcal{M}_{\text{flat}}(G_{\mathbb{C}},\Sigma(2,3,11))}n_{\alpha}e^{2\pi ik\text{CS}(\alpha)}Z_{\text{pert}}^{\alpha}.
\end{equation}
Furthermore, we have the transseries coefficients relating to the Stokes monodromy coefficients as specified:
\begin{equation*}
    n_{\alpha,0}=
    \begin{cases}
    1&\alpha=(\alpha_0,0)\\
    \frac{1}{2}m_{\alpha}^{\alpha_0,0}&\text{otherwise}.
    \end{cases}
    ,\qquad \xrightarrow{(practically)} \qquad 
    n_{\alpha}=
    \begin{cases}
    1&\text{if }\alpha\in\mathcal{M}_{\text{flat}}(SU(2),\Sigma(2,3,11))\\
    0&\text{if }\alpha\not\in\mathcal{M}_{\text{flat}}(SU(2),\Sigma(2,3,11)).
    \end{cases}
\end{equation*}
with 
\begin{equation*}
    m_{\beta}^{\alpha_0,0}=
    \begin{cases}
    -1,&\beta=(\alpha_1,-n^2/264), n=67,89,109,131\text{ mod }132,\\
    +1,&\beta=(\alpha_1,-n^2/264), n=1,23,43,65\text{ mod }132,\\
    -1,&\beta=(\alpha_2,-n^2/264), n=5,61,83,115\text{ mod }132,\\
    +1,&\beta=(\alpha_2,-n^2/264), n=17,49,71,127\text{ mod }132,\\
    -1,&\beta=(\alpha_3,-n^2/264), n=7,29,37,59\text{ mod }132,\\
    +1,&\beta=(\alpha_3,-n^2/264), n=73,95,103,125\text{ mod }132,\\
    -1,&\beta=(\alpha_4,-n^2/264), n=13,31,35,53\text{ mod }132,\\
    +1,&\beta=(\alpha_4,-n^2/264), n=79,97,101,119\text{ mod }132,\\
    -1,&\beta=(\alpha_5,-n^2/264), n=85,91,107,113\text{ mod }132,\\
    +1,&\beta=(\alpha_5,-n^2/264), n=19,25,41,47\text{ mod }132,\\
    0,&\text{otherwise.}
    \end{cases}
\end{equation*}
And at last, the remaining flat connection, $\alpha_2$ is also non-abelian, but can only be conjugated inside SL$(2,\mathbb{R})$ and is hence "complex". 

These residues of the poles determine the weights of these contributions to the integral in the generalized Borel resummation, and therefore the Chern-Simons partition function. Thus, the first group of poles with $a=1,23,43,65$ contributes
\begin{multline}
    \frac{i}{33}\left[\cos\left(\frac{8\pi}{33}\right)-\sin\left(\frac{5\pi}{66}\right)\right]\\
    \times \left(\sum_{n=\pm1\text{ mod }132}^{\infty}\pm 1+\sum_{n=\pm23\text{ mod }132}^{\infty}\pm 1
    +\sum_{n=\pm43\text{ mod }132}^{\infty}\pm1+\sum_{n=\pm 65\text{ mod }132}^{\infty}\pm1\right) e^{-\frac{1}{264}2\pi ik}\\
    =\frac{i}{33}\left[\cos\left(\frac{8\pi}{33}\right)-\sin\left(\frac{5\pi}{66}\right)\right]\sum_{a=1,23,43,65}\left(1-\frac{a}{66}\right)e^{-\frac{\pi ik}{132}}
\end{multline}
Thus, making
\begin{equation}
n_{\alpha_1}e^{2\pi ik\text{CS}(\alpha_1)}Z^{\alpha_1}_{\text{pert}}=\frac{2i}{33}\left[\cos\left(\frac{8\pi}{33}\right)-\sin\left(\frac{5\pi}{66}\right)\right]e^{-\frac{\pi ik}{132}}
\end{equation}
Performing the same computation gives us:
\begin{itemize}
    \item $\alpha_2$: 0 as $n_{\alpha_2}=0$
    \item $\alpha_3$:
    \begin{equation}
        n_{\alpha_3}e^{2\pi ik\text{CS}(\alpha_3)}Z^{\alpha_3}_{\text{pert}}=-\frac{2i}{33}\left[\cos\left(\frac{\pi}{33}\right)+\sin\left(\frac{13\pi}{66}\right)\right]e^{-\frac{49\pi ik}{132}}
    \end{equation}
    \item $\alpha_4$:
    \begin{equation}
        n_{\alpha_4}e^{2\pi ik\text{CS}(\alpha_4)}Z^{\alpha_4}_{\text{pert}}=-\frac{2i}{33}\left[\cos\left(\frac{5\pi}{33}\right)+\sin\left(\frac{\pi}{66}\right)\right]e^{-\frac{169\pi ik}{132}}
    \end{equation}
    \item $\alpha_5$:
    \begin{equation}
        n_{\alpha_5}e^{2\pi ik\text{CS}(\alpha_5)}Z^{\alpha_5}_{\text{pert}}=\frac{2i}{33}\left[\cos\left(\frac{2\pi}{33}\right)+\sin\left(\frac{7\pi}{66}\right)\right]e^{-\frac{361\pi ik}{132}}
    \end{equation}
\end{itemize}
From previous conjectures, we have 
\begin{equation}
\label{eq:result2311}
    Z_b(q)=\sum_{b}S_{ab}\hat{Z}_b(q)
\end{equation}
where the S-matrix is conjectured to exist and be k-independent (and $\hat{Z}_b(q)$ is our computed homological blocks) of which we sum over the abelian flat connections. Since we only have one abelian flat connection, $\alpha_0$, we have $\hat{Z}_{\alpha_0}\in q^{\Delta_{\alpha_0}}\mathbb{Z}[[q]]$ of which from our previous computations, we can then immediately identify $\Delta_{\alpha_0}=25/264$ and $S_{\alpha_0\alpha_0}=-3/2$.

\section{Conclusion}
Our main result Eq. \eqref{eq:result} provides a quick algorithm for performing the resurgent analysis of the SU(2) Chern-Simons partition function. Furthermore, from our computations done for $\Sigma(2,3,11)$, we observe that the results found in Eq. \eqref{eq:result2311} for $\Sigma(2,3,11)$ is presented as a special case of our general results Eq. \eqref{eq:result} when $n=1$. These results for the family of Brieskorn spheres provide further support for the construction and resurgent analysis of the homological blocks for plumbed manifolds proposed in \cite{GPV,GMP,GPPV}.

While we have been able to provide a general description that quickly reveals the non-perturbative properties of the SU(2) Chern-Simons partition function, we have not been able to generalize our results to include all Brieskorn spheres $\Sigma(2,3,6n+5),\ \forall n\in\mathbb{Z}_+$. In particular, during the course of our investigation, we have noticed missing flat connections for non-prime $6n+5$ Brieskorn spheres. This is a possible indication of the presence of renormalon effects, which has been discussed in \cite{MR,Gukov}, in our theory of which requires modification of our resurgent analysis done for the homological blocks and requires additional investigation on the topic.

\section*{Acknowledgement}
The author would like to thank Sergei Gukov for his invaluable comments and discussions during the project.

\end{document}